\documentclass[12pt]{article}
\usepackage{amssymb,amsthm,amsfonts}
 \usepackage{graphicx}
\usepackage[nonamelimits]{amsmath}
\usepackage{fullpage}

\newtheorem{theorem}{Theorem}

\newtheorem{claim}[theorem]{Claim}

\newtheorem{definition}[theorem]{Definition}

\begin{document}
\bibliographystyle{stdbib}

\title{~~\\[-1in]
%\\ [-1.75in]{\normalsize \hfill \tt Preliminary and Incomplete}\\
%[-1ex] {\normalsize \hfill \tt Please Do Not Quote}\\ [-1ex] {\normalsize
%\hfill \tt Comments Welcome}\\ [.125in]%
{A Computational View of Market Efficiency\thanks{The views and opinions expressed in this article
are those of the authors only, and do not necessarily represent the views
and opinions of AlphaSimplex Group, MIT, Northeastern University, any of
their affiliates and employees, or any of the individuals acknowledged
below. The authors make no representations or warranty, either expressed
or implied, as to the accuracy or completeness of the information
contained in this article, nor are they recommending that this article
serve as the basis for any investment decision---this article is for
information purposes only. This research was supported by the MIT
Laboratory for Financial Engineering and AlphaSimplex Group, LLC.}}}

\author{\\[.125in]Jasmina Hasanhodzic\thanks{Senior Research
Scientist, AlphaSimplex Group.}\ ,{\ } Andrew W. Lo\thanks{Harris
\& Harris Group Professor, MIT Sloan School of Management.}\ ,
\setcounter{footnote}{7}\ and Emanuele Viola\thanks{Assistant
Professor, College of Computer and Information Science, Northeastern
University, 440 Huntington Ave, \#246WVH, Boston, MA 02115, USA
(corresponding author).}}

%\date{First Draft: January 30, 2009 \\ [.125in]
%This Draft: February 4, 2009}

%%\date{February 8, 2009}

\maketitle \thispagestyle{empty} \centerline{\large \bf Abstract}
\baselineskip 14pt \vskip 20pt \noindent

We propose to study market efficiency from a
computational viewpoint. Borrowing from theoretical
computer science, we define a market to be
\emph{efficient with respect to resources $S$}
(e.g., time, memory) if no strategy using resources
$S$ can make a profit. As a first step, we consider
memory-$m$ strategies whose action at time $t$ depends only on
the $m$ previous observations at times
$t-m,\ldots,t-1$. We introduce and study a simple model of market
evolution, where strategies
impact the market by their decision to buy or sell.
We show that the effect of optimal strategies using
memory $m$ can lead to ``market conditions''
that were not present initially, such as (1) market bubbles and
(2) the possibility for a strategy using memory $m' > m$ to make a bigger profit than was initially
possible. We suggest ours as a framework to rationalize the technological arms race
of quantitative trading firms.


\thispagestyle{empty}
%\newpage

\section{Introduction}

Market efficiency---the idea that "prices fully reflect all
available information"---is one of the most important
concepts in economics. A large number of articles
have been devoted to its formulation, statistical
implementation, and refutation since Samuelson (1965) and
Fama (1965a,b; 1970) first argued that price changes must be
unforecastable if they fully incorporate the information
and expectations of all market participants.
The more efficient the market, the more random the
sequence of price changes generated by it, and the most
efficient market of all is one in which price changes are
completely random and unpredictable.

According to the proponents of market efficiency, this randomness is a direct result of many active market participants attempting to profit from their
information. Driven by profit opportunities, legions of
investors pounce on even the smallest informational
advantages at their disposal, and in doing so, they
incorporate their information into market prices and
quickly eliminate the profit opportunities that first
motivated their trades. If this occurs instantaneously,
as in an idealized world of "frictionless"
markets and costless trading, then prices
fully reflect all available information. Therefore, no
profits can be garnered from information-based trading
because such profits must have already been captured. In
mathematical terms, prices follow martingales.

This stands in sharp contrast to finance practitioners who
attempt to forecast future prices based on past ones. And
perhaps surprisingly, some of them do appear to make
consistent profits that cannot be attributed to chance
alone.

\subsection{Our Contribution}

In this paper we suggest that a reinterpretation of market
efficiency in computational terms might be the key to
reconciling this theory with the possibility of making
profits based on past prices alone. We believe that it does not make sense to talk about
market efficiency without taking into account that market
participants have bounded resources. In other words,
instead of saying that a market is ``efficient'' we should say, borrowing from theoretical computer
science, that a market is \emph{efficient
with respect to resources $S$}, e.g., time, memory, etc., \emph{if
no strategy using resources $S$ can generate a substantial
profit}. Similarly, we cannot say that investors act
optimally given all the available information, but rather
they act optimally within their resources. This allows for
markets to be efficient for some investors, but not for
others; for example, a computationally powerful hedge fund
may extract profits from a market which looks very
efficient from the point of view of a day-trader who has
less resources at his disposal---arguably the status
quo.

As is well-known, suggestions in this same spirit have already been made in the literature. For example, Simon (1955) argued that agents are not rational but \emph{boundedly rational}, which can be interpreted in modern terms as bounded in computational resources. Many other works in this direction are discussed in \S\ref{s-relatedwork}.
The main difference between this line of research and ours is: while it appears that most previous works use sophisticated continuous-time models, in particular explicitly addressing the market-making mechanism (which sets the price given agents' actions), our model is simple, discrete, and abstracts from the market-making mechanism.

\begin{figure}

\begin{center}
\includegraphics[trim = 30mm 10mm 0mm 10mm, clip, scale=0.45]{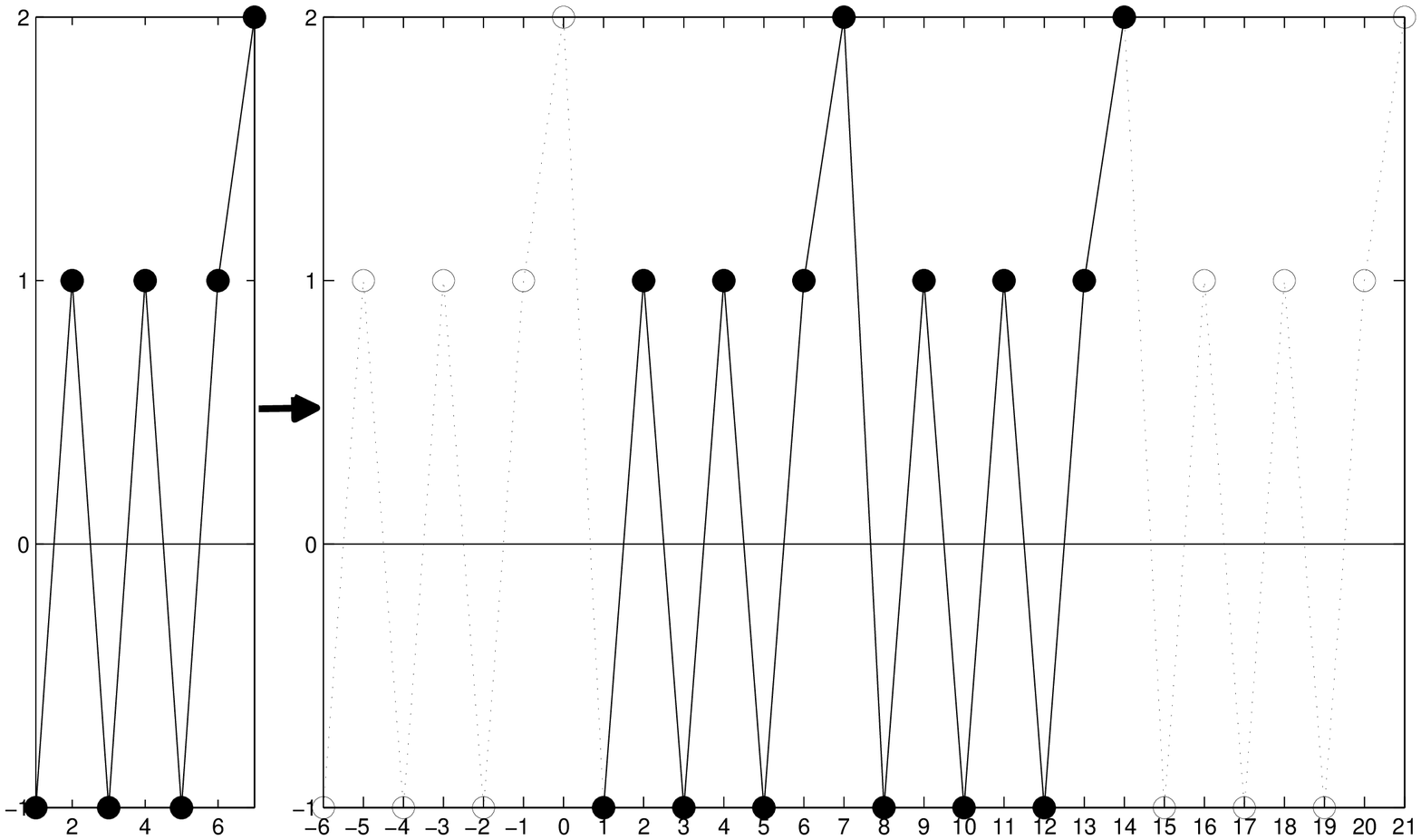}
\caption{The market corresponding to a market
pattern.} \label{figure:patternVsMarket}
\end{center}

\begin{center}
\includegraphics[trim = 30mm 10mm 0mm 10mm, clip, scale=0.45]{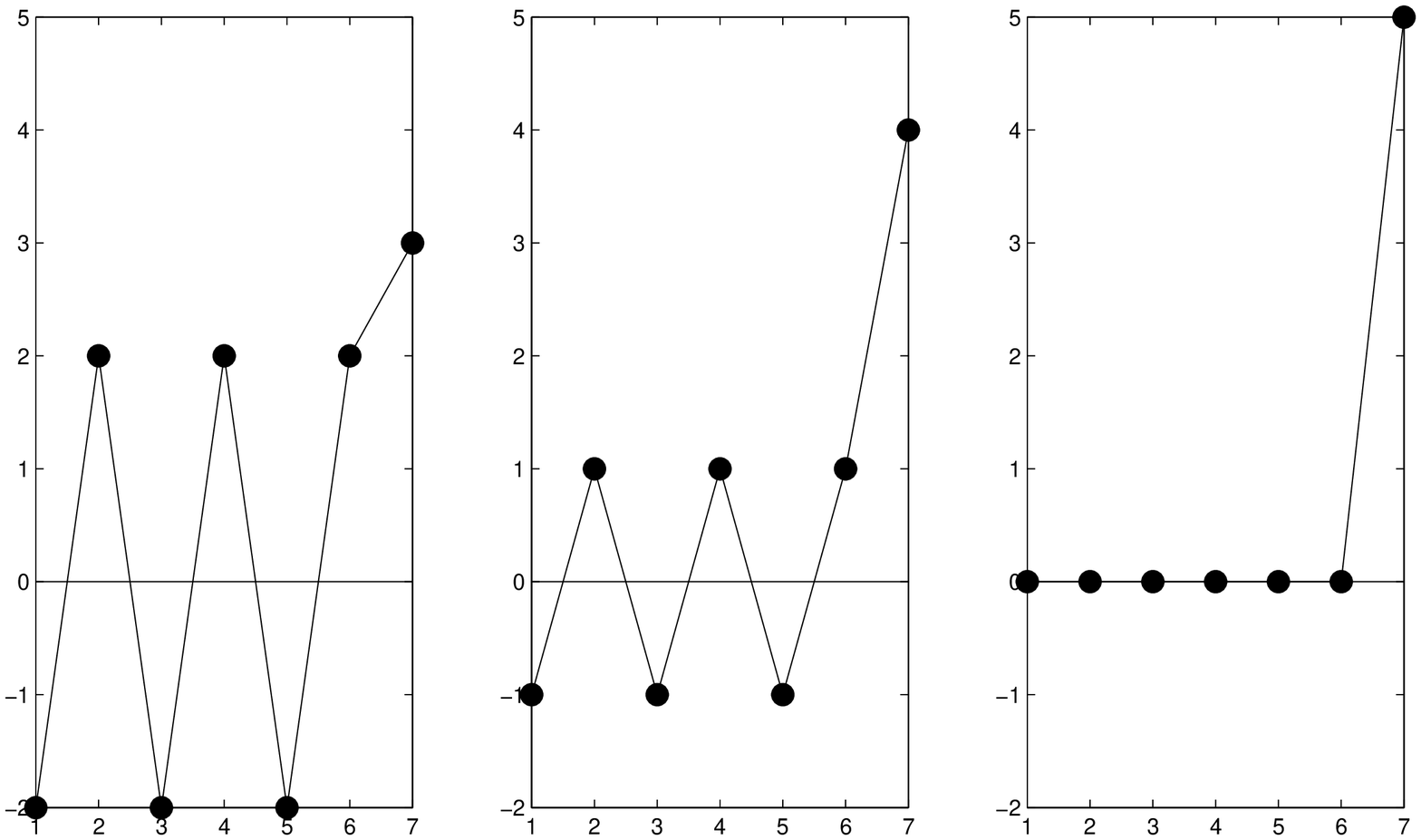}
\caption{Optimal memory-$2$ strategy evolves the
market and creates a bubble.} \label{figure:evolutionPlot}
\end{center}

\begin{center}
\includegraphics[trim = 30mm 10mm 0mm 10mm, clip, scale=0.45]{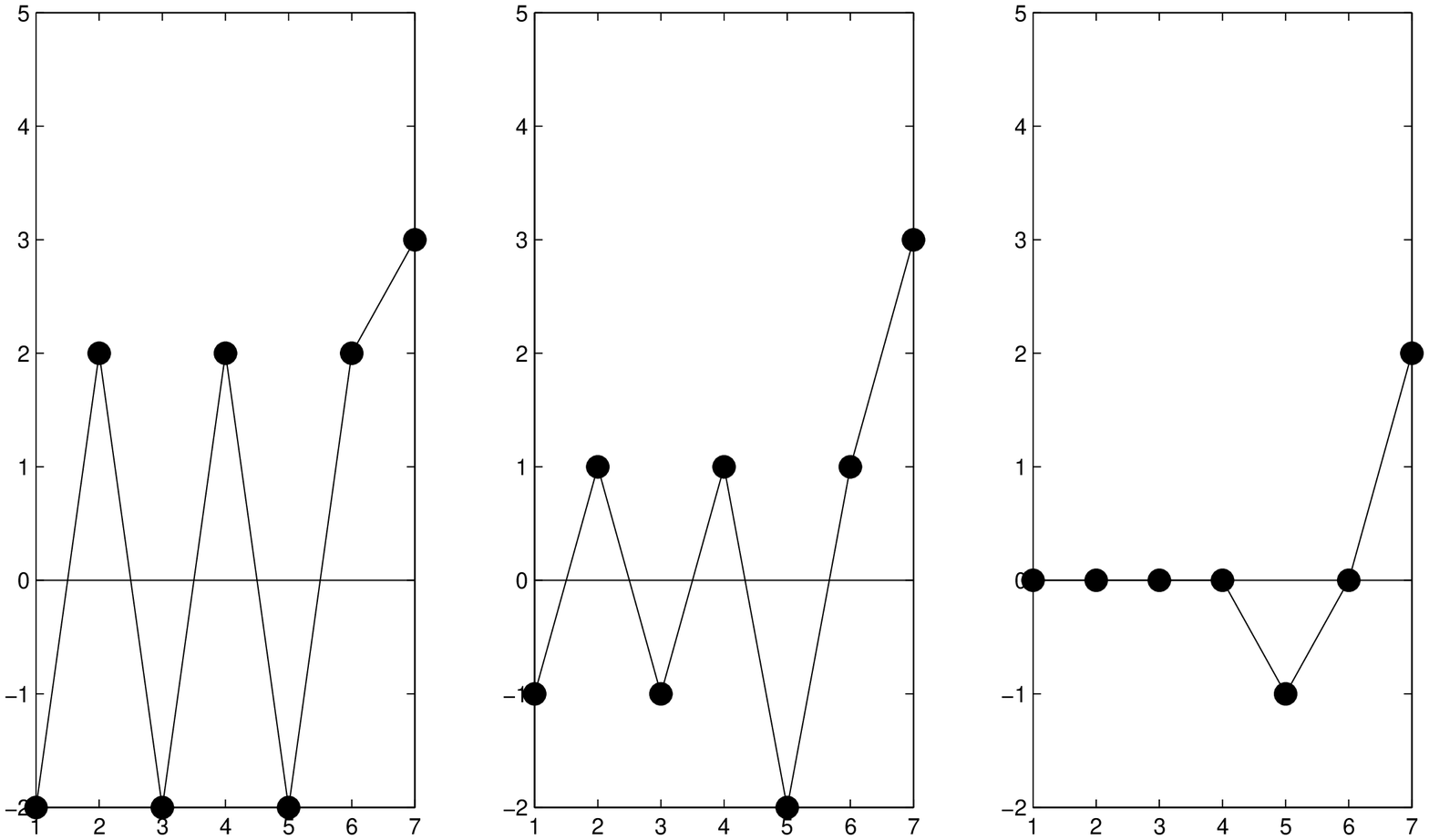}
\caption{Optimal memory-$3$ strategy evolves the
market without creating a bubble.} \label{figure:evolutionPlotmem3}
\end{center}
\end{figure}

\paragraph{Our model and results:}
We consider an economy where there exists only one good. The daily returns (price-differences) of this good follow a pattern, e.g.:
$$
(-1,1,-1,1,-1,1,2,-1,1,-1,1,-1,1,2,-1,1,-1,1,-1,1,2,\ldots),$$
which we write as $[-1,1,-1,1,-1,1,2]$, see Figure \ref{figure:patternVsMarket} on Page \pageref{figure:patternVsMarket}.
A good whose returns follow a pattern may arise from a variety of causes ranging from biological to political; think of seasonal cycles for commodities, or the 4-year presidential cycle. Though one can consider more complicated dependencies, working with these finite patterns keeps the mathematics simple and is sufficient for the points made below.

Now we consider a new agent $A$ who is allowed to trade the good daily, with no transaction costs. Intuitively, agent $A$ can profit whenever she can predict the sign of the next return. However, the key point is that $A$ is computationally bounded: her prediction can only depend on the previous $m$ returns. Continuing the above example and setting memory $m=2$, we see that upon seeing returns $(-1,1)$, $A$'s best strategy is to guess that the next return will be negative: this will be correct for $2$ out of the $3$ occurrences of $(-1,1)$ (in each occurrence of the pattern).

We now consider a model of market evolution where
$A$'s strategy impacts the market by pushing the
return closer to $0$ whenever it correctly forecasts the
sign of the return --- thereby exploiting the existing possibility
of profit --- or pushing it away from $0$ when the
forecast is incorrect --- thereby creating a new
possibility for profit. In other words, the evolved market is the difference between the original market and the strategy guess (we think of the guess in $\{-1,0,1\}$). For example, Figure \ref{figure:evolutionPlot} shows how the most profitable strategy using memory $m = 2$ evolves the market given by the pattern $[-2,2,-2,2,-2,2,3]$ (similar to the previous example).

\medskip

The main question addressed in this work is: \textbf{how do markets evolved by memory-bounded strategies look like?}

We show that the effect of optimal strategies using memory $m$ can lead to "market conditions" that were not present
initially, such as (1) market bubbles and (2) the
possibility that a strategy using memory $m' > m$ can
make larger profits than previously possible.
By market bubbles (1), we mean that some returns will grow much higher, an effect already anticipated in Figure \ref{figure:evolutionPlot}.
For the point (2), we consider a new agent $B$ that has memory $m'$ that is larger than that of $A$. We let $B$ trade on the market evolved by $A$. We show that, for some initial market, $B$ may make more profit than what would have been possible if $A$ had not evolved the market, or even if another agent with large memory $m'$ had evolved the market instead of $A$. Thus, it is precisely the presence of low-memory agents ($A$) that allows high-memory agents ($B$) to make large profits.

Regarding the framework for (2), we stress that we only consider agents trading sequentially, not simultaneously: after $A$ evolves the market, it gets incorporated into a new market, on which $B$ may trade. While a natural direction is extending our model to simultaneous agents, we argue that this sequentiality is not unrealistic. Nowadays, some popular strategies $A$ are updated only once a month. Within any such month, a higher-frequency strategy $B$ can indeed trade in the market evolved by $A$ without $A$ making any adjustment.

\subsection{More related work} \label{s-relatedwork}

Since Samuelson's (1965) and Fama's (1965a,b; 1970) landmark papers, many others extended their original framework, yielding a ``neoclassical''
version of the efficient market hypothesis where price changes, properly weighted by aggregate
marginal utilities, are unforecastable (see, for example, LeRoy, 1973;
Rubinstein, 1976; and Lucas, 1978).  In markets where, according to Lucas
(1978), all investors have ``rational expectations,'' prices do fully
reflect all available information and marginal-utility-weighted prices
follow martingales. Market efficiency has been extended in many other
directions, but the general thrust is
the same: individual investors form expectations rationally, markets
aggregate information efficiently, and equilibrium prices incorporate all available information instantaneously. See Lo (1997, 2007) for a more detailed summary of the market efficiency literature in economics and finance.

There are two branches of the market efficiency literature that are
particularly relevant for our paper: the asymmetric information
literature, and the literature on asset bubbles and crashes.  In Fischer Black's (1986) presidential address to the American Finance Association,
he argued that financial market prices were subject to ``noise'', which
could temporarily create inefficiencies that would ultimately be
eliminated through intelligent investors competing against each other to
generate profitable trades.  Since then, many authors have modeled
financial markets by hypothesizing two types of traders---informed and
uninformed---where informed traders have private information regarding the
true economic value of a security, and uninformed traders have no
information at all, but merely trade for liquidity needs (see, for
example, Grossman and Stiglitz, 1980; Diamond and Verrecchia, 1981;
Admati, 1985; Kyle, 1985; and Campbell and Kyle, 1993). In this context,
Grossman and Stiglitz (1980) argue that market efficiency is impossible
because if markets were truly efficient, there would be no incentive for
investors to gather private information and trade.  DeLong et al.\ (1990,
1991) provide a more detailed analysis in which certain types of
uninformed traders can destabilize market prices for periods of time even
in the presence of informed traders.  More recently, studies by Luo (1995,
1998, 2001, 2003), Hirshleifer and Luo (2001), and Kogan et al. (2006)
have focused on the long-term viability of noise traders when competing
for survival against informed traders; while noise traders are exploited
by informed traders as expected, certain conditions do allow them to
persist, at least in limited numbers.

The second relevant offshoot of the market efficiency literature involves
``rational bubbles'', in which financial asset prices can become
arbitrarily large even when all agents are acting rationally (see, for
example, Blanchard and Watson, 1982; Allen, Morris, and Postlewaite,
1993; Santos and Woodford, 1997; and Abreu and Brunnermeier, 2003).  LeRoy
(2004) provides a comprehensive review of this literature.

Finally, in the computer science literature, market efficiency has yet to
be addressed explicitly.  However, a number of studies have touched upon
this concept tangentially.  For example, computational learning methods
have been applied to the pricing and trading of financial securities by
Hutchinson, Lo, and Poggio (1994). Evolutionary
approaches in the dynamical systems and complexity literature have also
been applied to financial markets by Farmer and Lo (1999), Farmer (2002),
Farmer and Joshi (2002), Lo (2004), Farmer, Patelli, and Zovko (2005), Lo
(2005), and Bouchaud, Farmer, and Lillo (2008).  And
simulations of market dynamics using autonomous agents have coalesced into
a distinct literature on ``agent-based models'' of financial markets by
Arthur et al. (1994, 1997), Arthur, LeBaron, and Palmer (1999), Farmer
(2001), and LeBaron (2001a--c, 2002, 2006). While none of the papers
in these distinct strands of the computational markets literature focus
on a new definition of market efficiency, nevertheless, they all touch
upon different characteristics of this aspect of financial markets.

\iffalse

We also point out that there are many works in computer science that have contributed an analytical study of \emph{game theory} with computationally bounded agents, see for example Halpern and Pass (2008) and the references therein. Here, typically, computationally bounded agents are modeled via finite automata. We are unaware of any work in this strand addressing market efficiency.

\fi

\paragraph{Organization.} We provide our
computational definition of market efficiency in Section
$\ref{s-compEMH}$. We describe the dynamics of market evolution in
Section $\ref{s-evolution}$, and conclude in Section $\ref{s-conclusion}$.

\section{A computational definition of market efficiency} \label{s-compEMH}

We model a market by an infinite sequence of random
variables $X_1,X_2,\ldots$ where $X_t \in
\mathbb{R}$ is the \emph{return} at time $t$. For
the points made in this work, it is enough to
consider markets that are obtained by the
repetition of a pattern. This has the benefit of
keeping the mathematics to a minimum level of
sophistication.

\begin{definition}[Market]
A \emph{market pattern} of length $p$ is a sequence
of $p$ random variables $[X_1,\ldots,X_p]$, where
each $X_i \in \mathbb{R}$.
\end{definition}

A market pattern $[X_1,\ldots,X_p]$ gives rise to a
market obtained by the \emph{independent}
repetition of the patterns:
$X^1_1,\ldots,X^1_p,X^2_1,\ldots,X^2_p,X^3_1,\ldots,X^3_p,\ldots$,
where each block of random variables
$X^i_1,\ldots,X^i_p$ are distributed like
$X_1,\ldots,X_p$ and independent from
$X^j_1,\ldots,X^j_p$ for $j \ne i$. Figure
\ref{figure:patternVsMarket} shows an example for a
market pattern in which the random variables are
constant.

We now define strategies. A memory-$m$ strategy takes as input
the previous $m$ observations of market returns,
and outputs $1$ if it thinks that the next return
will be positive. We can think of this ``$1$'' as
corresponding to a buy-and-sell order (which is
profitable if the next return is indeed positive).
Similarly, a strategy output of $-1$ corresponds to
a sell-and-buy order, while $0$ corresponds to no
order.

\begin{definition}
A \emph{memory-$m$ strategy} is a map $s :
\mathbb{R}^m \to \{-1,0,1\}$.
\end{definition}

\newcommand{\tcost}{\ensuremath{\mathit{TCost}}}
\newcommand{\sign}{\ensuremath{\mathit{sign}}}

%%%explain pattern vs. market

We now define the gain of a strategy over a market.
At every market observation, the gain of the strategy
is the sign of the
product of the strategy output and the market
return: the strategy gains a profit when it correctly predicts whether the next return will be positive or negative, and loses a profit otherwise.
 Since we think of markets as defined
by the repetition of a pattern, it is enough
to define the gain of the strategy over this pattern.

\begin{definition}[Gain of a strategy]
The \emph{gain} of a memory-$m$ strategy $s :
\mathbb{R}^m \to \{-1,0,1\}$ over market pattern
$[X_1,\ldots,X_p]$ is:
$$ \sum_{i = 1}^p E_{X_1,\ldots,X_p}\left[
\sign\left(s(X_{i-m},X_{i-m+1},\ldots,X_{i-1})
\cdot X_i \right)\right],$$
where for any $k$, the random variable $(X_{1 - k\cdot p},\ldots,X_{p - k\cdot p})$ is independent from all the others and distributed like $(X_1,\ldots,X_p)$.

A memory-$m$ strategy $s$ is \emph{optimal} over a
market pattern if no memory-$m$ strategy has a
bigger gain than $s$ over that market pattern.
\end{definition}

The gain of an optimal memory-$1$ strategy over a certain market is a concept which is intuitively related to the (first-order) \emph{autocorrelation} of the market, i.e.~the correlation between the return at time $t$ and that at time $t+1$ (for random $t$), a well-studied measure of market efficiency (cf.~Lo 2005, figure 2). This analogy can be made exact up to a normalization for markets that are given by balanced sequences of $+1,-1$. In this case higher memory can be thought of an extension of autocorrelation. It may be interesting to apply this measure to actual data.

Given a deterministic market pattern (which is just a sequence
of numbers such as $[2,-1,5,...])$, optimal strategies can be easily computed: it is easy to see that an optimal strategy outputs $+1$ if and only if more than
half the occurrences of $x$ in the market pattern are followed by a positive value.

We are now ready to give our definition of market efficiency.

\begin{definition}[Market efficiency]
A market pattern $[X_1,\ldots,X_p]$ is
\emph{efficient} with respect to memory-$m$
strategies if no such strategy has strictly
positive gain over $[X_1,\ldots,X_p]$.
\end{definition}

For example, let $X_1$ be a $\{-1,1\}$ random
variable that is $-1$ with probability $1/2$. Then
the market pattern $[X_1]$ is efficient with
respect to memory-$m$ strategies for every $m$.

A standard ``parity'' argument gives the following
hierarchy, which we prove for the sake of completeness.

\begin{claim} For every $m$, there is a market
pattern that is efficient for memory-$m$
strategies, but is not efficient for memory-$(m+1)$
strategies.
\end{claim}
\begin{proof}
Let $X_1,X_2,\ldots,X_{m+1}$ be i.i.d.~random
variables with range $\{-1,1\}$ whose probability
of being $1$ is $1/2$. Consider the market pattern
$$\left[X_1,X_2,\ldots,X_{m+1}, X_{m+2} := \prod_{i = 1}^{m+1} X_i\right]$$
of length $p = m+2$.
By the definition of the pattern, the distribution of each variable is independent from the previous $m$, and therefore any memory-$m$ strategy has gain $0$.

Consider the memory-$(m+1)$ strategy $s(x_1,x_2,\ldots,x_{m+1}) := \prod_{i = 1}^{m+1} x_i$. Its gain over the market pattern is
$$\sum_{i = 1}^p E_{X_1,\ldots,X_p}\left[
\sign\left(X_{i-m-1} \cdot X_{i-m} \cdots X_{i-1}
\cdot X_i \right)\right] =
E_{X_1,\ldots,X_p}\left[
\sign\left(X_{1} \cdot X_{2} \cdots X_{m+1}
\cdot X_{m+2} \right)\right] = 1.$$
\end{proof}

\section{Market evolution} \label{s-evolution}

In this section we describe the dynamics of our market model.
We consider a
simple model of evolution where the strategies
enter the market sequentially. After a strategy
enters, its ``impact on the market'' is recorded in the market and
produces a new evolved market. The way in which a strategy impacts the market is by subtracting the strategy output from the market data.

\begin{definition}[Market evolution] \label{def-marketevolution}
We say that a memory-$m$ strategy $s : \mathbb{R}^m
\to \{-1,0,1\}$ \emph{evolves} a market pattern
$[X_1,X_2,\ldots,X_p]$ into the market pattern
$$[X_1 - s(X_{1-m},\ldots,X_{1-1}), X_2 -
s(X_{2-m},\ldots,X_{2-1}), \ldots, X_p -
s(X_{p-m},\ldots,X_{p-1}) ].$$
\end{definition}

Figure \ref{figure:evolutionPlot} shows the
evolution of a market pattern.

Definition \ref{def-marketevolution} can be readily extended to multiple strategies acting simultaneously, just by subtracting all the impacts of the strategies on the market, but for the points made in this paper the above one is enough.

In the next two subsections we point out two consequences of the above definition.

\subsection{Market bubbles}

\begin{figure}

\begin{center}
\includegraphics[trim = 30mm 10mm 0mm 10mm, clip, scale=0.45]{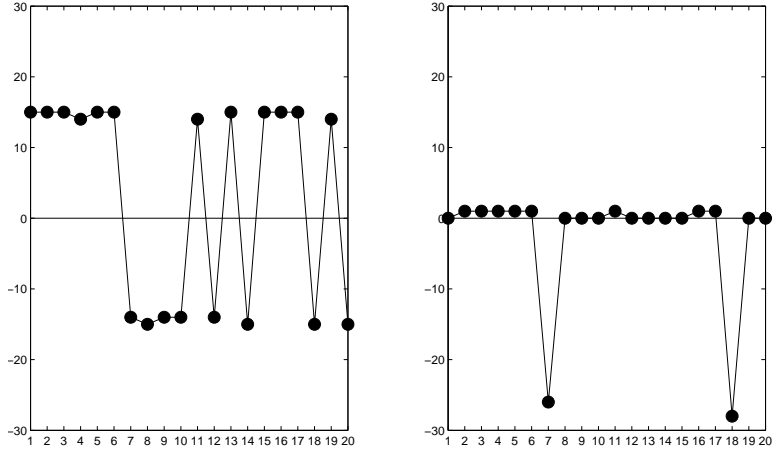}
\caption{Random market pattern (on the left) evolved by an optimal memory $3$ strategy into a bubble after $16$ iterations (on the right).} \label{figure:bubbleFromRandom}
\end{center}

\end{figure}

In this section we point out how low-memory strategies can give rise to \emph{market bubbles}, i.e.~we show that an optimal low-memory strategy can evolve a market pattern into another one which has values that are much bigger than those of the original market pattern (in absolute value). Consider for example the market pattern $[-2,2,-2,2,-2,2,3]$ in Figure \ref{figure:evolutionPlot}. Note how an
\emph{optimal} memory-$2$ strategy will profit on
most time instances but not all. In particular, on input $(-2,2)$, the best output is $-1$, which agrees with the sign of the market in $2$ out of $3$ occurrences of $(-2,2)$. This shrinks the
returns towards $0$ on most time instances, but
will make the last return in the pattern rise.
The optimal memory-$2$ strategy evolves the original market pattern into $[-1,1,-1,1,-1,1,3]$. The situation then repeats, and an optimal memory-$2$ strategy evolves the latter pattern into $[0,0,0,0,0,0,5]$. This is an example of how an optimal strategy creates market conditions which were not initially present.

We point out that the market bubble does not form with memory $3$; see Figure \ref{figure:evolutionPlotmem3}.

We also point out that the formation of such bubbles is not an isolated phenomenon, nor specific to memory $2$: we have generated several random markets and plotted their evolutions, and bubbles often arise with various memories. We report one such example in Figure \ref{figure:bubbleFromRandom}, where a random pattern of length $20$ is evolved by optimal memory-$3$ strategies into a bubble after $16$ iterations.
In fact, the framework proposed in this paper gives rise to a ``game'' in the spirit of Conway's Game of Life, showing how simple trading rules can lead to apparently chaotic market dynamics. Perhaps the main difference between our game and Conway's is that in ours strategies perform optimally within their resources.

\subsection{High-memory strategies feed off low-memory ones}

The next claim shows a pattern where a high-memory
strategy can make a bigger profit after a
low-memory strategy has acted and modified the
market pattern. This profit is bigger than the
profit that is obtainable by a high-memory
strategy without the low-memory strategy acting
beforehand, and even bigger than the profit
obtainable after another high-memory strategy acts
beforehand. Thus it is precisely the presence of
low-memory strategies that creates opportunities
for high-memory strategies which were not present
initially. This example provides explanation for
the real-life status quo which sees a growing
quantitative sophistication among asset managers.

Informally, the proof of the claim exhibits a market with a certain ``symmetry.'' For high-memory strategies, the best choice is to maintain the symmetry by profiting in multiple points. But a low-memory strategy will be unable to do. Its optimal choice will be to ``break the symmetry,'' creating new profit opportunities for high-memory strategies.

\begin{claim} \label{claim-feedoff}
For every $m < m'$ there are a market pattern $P =
[X_1,\ldots,X_p]$, an optimal memory-$m$ strategy $s_m$ that evolves $P$ into $P_m$, and an optimal memory-$m'$ strategy $s_{m'}$ that evolves $P$ into $P_{m'}$ such that the gain of an optimal memory-$m'$ strategy over
$P_{m}$ is bigger than either of the following:
\begin{itemize}
\item the gain of $s_{m'}$ over $P$,
\item the gain of any memory-$m$ strategy over $P_{m}$,
\item the gain of any memory-$m'$ strategy over $P_{m'}$.
\end{itemize}
\end{claim}
\begin{proof}
Consider the market pattern
\begin{align*}
& \Big[X_1,\ldots,X_{m-1}, a \cdot X_m, b \prod_{i = 1}^m X_i, Y_2, \ldots, Y_{m'-1},c \cdot Y_{m'}, \prod_{i =1}^m X_i \prod_{i = 2}^{m'} Y_i,\\
& X_1,\ldots,X_{m'-1}, a' \cdot X_{m'}, b \prod_{i = 1}^{m'} X_i , Y_2, \ldots, Y_{m'-1},c \cdot Y_{m'}, -\prod_{i =1}^{m'} X_i \prod_{i = 2}^{m'} Y_i \Big],
\end{align*}
where $X_1,\ldots,X_{m'}$ and $Y_2, \ldots, Y_{m'}$ are independent $\{-1,1\}$ random variables (note some of these variables appear multiple times in the pattern), and $a = 10, a' = 20, b = 30, c = 40$.

We now analyze the gains of various strategies. For this it is convenient to define the \emph{latest input} of a strategy $s(x_1,\ldots,x_{\ell})$ as $x_{\ell}$; this corresponds to the most recent observation the strategy is taking into consideration.

\emph{The strategy $s_m$:} We note that there is an optimal strategy $s_m$ on $P$ that outputs $0$ unless its latest input has absolute valute $a$. This is because in all other instances the random variable the strategy is trying to predict is independent from the previous $m$. Thus, $s_m(x_1,\ldots,x_m)$ equals $0$ unless $|x_m| = a$ in which case it outputs the sign of $\prod_{i=1}^m x_i$. The gain of this strategy is $1$. The strategy evolves the pattern into the same pattern except that the {\bfseries first} occurrence of $b$ is replaced with $b-1$:
\begin{align*}
P_m := & \Big[X_1,\ldots,X_{m-1}, a \cdot X_m, (b-1) \prod_{i = 1}^m X_i, Y_2, \ldots, Y_{m'-1},c \cdot Y_{m'}, \prod_{i =1}^m X_i \prod_{i = 2}^{m'} Y_i,\\
& X_1,\ldots,X_{m'-1}, a' \cdot X_{m'}, b \prod_{i = 1}^{m'} X_i , Y_2, \ldots, Y_{m'-1},c \cdot Y_{m'}, -\prod_{i =1}^{m'} X_i \prod_{i = 2}^{m'} Y_i \Big].
\end{align*}

\emph{The strategy $s_{m'}$:} We note that there is an optimal memory-$m'$ strategy over $P$ that outputs $0$ unless its latest input has absolute value $a,a',$ or $c$. Again, this is because in all other instances the strategy is trying to predict a variable that is independent from the previous $m'$. Moreover, when its latest input has absolute value $c$, we can also assume that the strategy outputs $0$. This is because the corresponding contribution is
$$E\left[\sign\left(s_{m'}(b \cdot \prod_{i = 1}^m X_i, Y_2, \ldots, Y_{m'-1},c \cdot Y_{m'}) \cdot \prod_{i =1}^m X_i \prod_{i = 2}^{m'} Y_i\right)\right]$$
$$+ E\left[\sign\left(s_{m'}(b \cdot \prod_{i = 1}^{m'} X_i, Y_2, \ldots, Y_{m'-1},c \cdot Y_{m'}) \cdot - \prod_{i =1}^{m'} X_i \prod_{i = 2}^{m'} Y_i\right)\right] = 0.$$
Thus there is an optimal memory-$m'$ strategy with gain $2$ that evolves the market pattern into the pattern $P_{m'}$ that is like $P$ with $b$ replaced by $b-1$ in {\bfseries both} occurrences (as opposed to $P_m$ which has $b$ replaced by $b-1$ only in the first occurrence).

\emph{The optimal memory-$m'$ strategy on $P_{m'}$.} Since $P_{m'}$ is like $P$ with $b$ replaced by $b-1$, the gain of the optimal memory-$m'$ strategy on $P_{m'}$ is $2$.

\emph{The optimal memory-$m$ strategy on $P_{m}$.}
Essentially the same argument for $s_m$ can be applied again to argue that any memory-$m$ strategy on $P_m$ has gain $1$ at most.

\emph{The optimal memory-$m'$ strategy on $P_{m}$.} Finally, note that there is a memory-$m'$ strategy whose gain is $4$ on $P_m$. This is because the replacement of the first occurrence of $b$ in $P$ with $b-1$ allows a memory-$m'$ strategy to predict the sign of the market correctly when its latest input has absolute value $c$.
\end{proof}

We make few final comments regarding Claim \ref{claim-feedoff}. First, the same example can be obtained for deterministic market patterns by considering the exponentially longer market pattern where the random variables take all possible combinations. However the randomized example is easier to describe.
Also, although in Claim \ref{claim-feedoff} we talk about \emph{a} optimal strategy, in fact the particular strategies considered there are natural in that they are those that minimize the number of $\{-1,1\}$ outputs (i.e., the amount of trading). We could have forced this to be the case by adding in our definition of gain a negligible transaction cost, in which case the claim would talk about \emph{the} optimal strategy, but we preferred a simpler definition of gain.

\section{Conclusion} \label{s-conclusion}

In this work we have suggested to study market efficiency from a computational point of view. We have put forth a specific memory-based framework which is simple and tractable, yet capable of modeling market dynamics such as the formation of market bubbles and the possibility for a high-memory strategy to ``feed off'' low-memory ones. Our results may provide an analytical framework for studying
the technological arms race that portfolio managers have
been engaged in since the advent of organized financial
markets.

Our framework also gives rise to a few technical questions, such
as how many evolutions does it take a pattern to reach a certain other
pattern, to what extent does this number of evolutions depends on
different levels of memory, and to what extent does it depend on the
simultaneous interaction among strategies using different memories.

\bigskip

\Large \centerline{{\bf References}} \vskip 30pt \normalsize \baselineskip
12pt
\begin{description}
\item
Abreu, D. and M. Brunnermeier, 2003, ``Bubbles and Crashes'',
\textit{Econometrica} 71, 173--204.

\item
Admati, A., 1985, ``A Noisy Rational Expectations Equilibrium for
Multi-asset Securities Markets'', \textit{Econometrica} 53, 629--657.

\item
Allen, F., Morris, S. and A. Postlewaite, 1993, ``Finite Bubbles with
Short Sale Constraints and Asymmetric Information'', \textit{Journal
of Economic Theory} 61, 206--229.

\item
Arthur, B., LeBaron, B. and R. Palmer, 1999, ``The Time Series
Properties of an Artificial Stock Market'', \textit{Journal of
Economic Dynamics and Control} 23, 1487--1516.

\item
Arthur, B., Holland, J., LeBaron, B., Palmer, R. and P. Tayler, 1994,
``Artificial Economic Life: A Simple Model of a Stock Market'',
\textit{Physica D,} 75, 264--274.

\item
Arthur, B., Holland, J., LeBaron, B., Palmer, R. and P. Tayler, 1997,
``Asset Pricing Under Endogenous Expectations in an Artificial Stock
Market'',  in B. Arthur, S. Durlauf, and D. Lane, eds., \textit{The
Economy as an Evolving Complex System II}, Reading, MA:
Addison-Wesley.

\item
Black, F., 1986, ``Noise'', \textit{Journal of Finance} 41, 529--544.

\item
Blanchard, O. and M. Watson, 1982, ``Bubbles, Rational Expectations
and Financial Markets'', in P. Wachtel (ed.), \textit{Crisis in the
Economic and Financial Structure: Bubbles, Bursts, and Shocks}.
Lexington, MA: Lexington Press.

\item
Bouchaud, J., Farmer, D. and F. Lillo, 2008, ``How Markets Slowly
Digest Changes in Supply and Demand'', in T. Hens and K. Schenk-Hoppe,
eds., \textit{Handbook of Financial Markets: Dynamics and Evolution}.
Elsevier: Academic Press.

\item
Campbell, J. and A. Kyle,  1993, ``Smart Money, Noise Trading and
Stock Price Behavior'', \textit{Review of Economic Studies} 60, 1--34.

\item Chang, K., C. Osler, 1999. ``Methodical Madness: Technical Analysis and the Irrationality of Exchange-rate Forecasts''. \textit{The Economic Journal} 109, 636--661.

\item
DeLong, B., Shleifer, A., Summers, L. and M. Waldman, 1990, ``Noise
Trader Risk in Financial Markets'', \textit{Journal of Political
Economy} 98, 703--738.

\item
DeLong, B., Shleifer, A., Summers, L. and M. Waldman, 1991, ``The
Survival of Noise Traders in Financial Markets'', \textit{Journal of
Business} 64, 1--19.

\item
Diamond, D. and R. Verrecchia, 1981, ``Information Aggregation in a
Noisy Rational Expectations Economy'', \textit{Journal of Financial
Economics} 9, 221--235.

\item
Fama, E., 1965a, ``The Behavior of Stock Market Prices'', {\it Journal
of Business}, 38, 34--105.

\item
Fama, E., 1965b, ``Random Walks In Stock Market Prices'',
\textit{Financial Analysts Journal} 21, 55--59.

\item
Fama, E., 1970, ``Efficient Capital Markets: A Review of Theory and
Empirical Work'', \textit{Journal of Finance} 25, 383--417.

\item
Farmer, D., 2001, ``Toward Agent-Based Models for Investment'', in
\textit{Developments in Quantitative Investment Models}. Charlotte,
NC: CFA Institute.

\item
Farmer, D., 2002, ``Market Force, Ecology and Evolution'',
\textit{Industrial and Corporate Change} 11, 895--953.

\item
Farmer, D., Gerig, A., Lillo, F. and S. Mike, 2006, ``Market
Efficiency and the Long-Memory of Supply and Demand: Is Price Impact
Variable and Permanent or Fixed and Temporary?'', \textit{Quantitative
Finance} 6, 107--112.

\item
Farmer, D. and S. Joshi, 2002, ``The Price Dynamics of Common Trading
Strategies'', \textit{Journal of Economnic Behavior and Organizations}
49, 149--171.

\item
Farmer, D. and A. Lo, 1999, ``Frontiers of Finance: Evolution and
Efficient Markets'', \textit{Proceedings of the National Academy of
Sciences} 96, 9991--9992.

\item
Farmer, D., Patelli, P. and I. Zovko, 2005, ``The Predictive Power of
Zero Intelligence in Financial Markets'', \textit{Proceedings of the
National Academy of Sciences} 102, 2254--2259.

\item
Grossman, S. and J. Stiglitz, 1980, ``On the Impossibility of
Informationally Efficient Markets'', \textit{American Economic Review}
70, 393--408.

\iffalse
\item Halpern, J. and R. Pass 2008, ``Game Theory with Costly Computation'', \textit{arXiv.org:0809.0024}.

\fi

\item
Hirshleifer, D. and G. Luo, 2001, ``On the Survival of Overconfident
Traders in a Competitive Securities Market'', \textit{Journal of
Financial Markets} 4, 73--84.

\item
Hutchinson, J., Lo, A. and T. Poggio, 1994, ``A Nonparametric Approach
to Pricing and Hedging Derivative Securities via Learning Networks'',
\textit{Journal of Finance} 49, 851--889.

\item
Kogan, Leonid, Stephen A. Ross, Jiang Wang, and Mark M. Westerfield,
2006, The price impact and survival of irrational traders, Journal of
Finance 61, 195–229.

\item
Kyle, A., 1985, ``Continuous Auctions and Insider Trading'',
\textit{Econometrica} 53, 1315--1335.

\item
LeBaron, B., 2001a, ``Financial Market Efficiency in a Coevolutionary
Environment'', \textit{Proceedings of the Workshop on Simulation of
Social Agents: Architectures and Institutions}, Argonne National
Laboratory and The University of Chicago, 33-51.

\item
LeBaron, B., 2001b, ``A Builder's Guide to Agent-Based Financial
Markets'', \textit{Quantitative Finance} 1, 254-261.

\item
LeBaron, B., 2001c, ``Evolution and Time Horizons In An Agent-Based
Stock Market'', \textit{Macroeconomic Dynamics} 5, 225--254.

\item
LeBaron, B., 2002, ``Short-Memory Traders and Their Impact on Group
Learning in Financial Markets'', \textit{Proceedings of the National
Academy of Science} 99, 7201--7206.

\item
LeBaron, B., 2006, ``Agent-Based Computational Finance'', in L.
Tesfatsion and K. Judd, eds., \textit{Handbook of Computational
Economics}. Amsterdam: North-Holland.

\item LeRoy, S., 1973. ``Risk aversion and the martingale property of stock returns''. \emph{International
Economic Review} 14, 436–46.

\item
LeRoy, S., 2004, ``Rational Exuberance'', \textit{Journal of Economic
Literature} 42, 783--804.

\item
Lo, A., ed., 1997, \textit{Market Efficiency: Stock Market Behaviour
In Theory and Practice, Volumes I and II}.  Cheltenham, UK: Edward
Elgar Publishing Company.

\item
Lo, A., 2004, ``The Adaptive Markets Hypothesis: Market Efficiency
from an Evolutionary Perspective'', {\it Journal of Portfolio
Management} 30, 15--29.

\item
Lo, A., 2005, ``Reconciling Efficient Markets with Behavioral Finance:
The Adaptive Markets Hypothesis'', {\it Journal of Investment
Consulting} 7, 21--44.

\item
Lo, A., 2007, ``Efficient Markets Hypothesis'', in \textit{The New
Palgrave: A Dictionary of Economics}, 2nd Edition, 2007.

\item
Lo, A., 2008, \textit{Hedge Funds: An Analytic Perspective}.
Princeton, NJ: Princeton University Press.

\item
Lo, A. and C. MacKinlay, 1999, \textit{A Non-Random Walk Down Wall
Street}. Princeton, NJ: Princeton University Press.

\item Lo, A., Mamaysky H., Wang J., 2000, ``Foundations of Technical Analysis: Computational Algorithms, Statistical Inference, and Empirical Implementation'', \textit{The Journal of Finance}, Volume LV (4), 1705--1765.

\item
Lucas, R., 1978, ``Asset Prices in an Exchange Economy'', {\it
Econometrica} 46, 1429--1446.

\item
Luo, G., 1995, ``Evolution and Market Competition'', \textit{Journal
of Economic Theory} 67, 223--250.

\item
Luo, G., 1998, ``Market Efficiency and Natural Selection in a
Commodity Futures Market'', \textit{Review of Financial Studies} 11,
647--674.

\item
Luo, G., 2001, ``Natural Selection and Market Efficiency in a Futures
Market with Random Shocks'', \textit{Journal of Futures Markets} 21,
489--516,

\item
Luo, G., 2003, ``Evolution, Efficiency and Noise Traders in a
One-Sided Auction Market'', \textit{Journal of Financial Markets} 6,
163--197.

\item
Roberts, H., 1959, ``Stock-Market `Patterns' and Financial Analysis:
{M}ethodological Suggestions'', {\it Journal of Finance} 14, 1--10.

\item
Roberts, H., 1967, ``Statistical versus Clinical Prediction of the
Stock Market'', unpublished manuscript, Center for Research in
Security Prices, University of Chicago, May.

\item Rubinstein, M., 1976., ``The valuation of uncertain income streams and the pricing of options''. {\it
Bell Journal of Economics} 7, 407--25.

\item
Samuelson, P., 1965, ``Proof that Properly Anticipated Prices
Fluctuate Randomly'', {\it Industrial Management Review} 6, 41--49.

\item
Santos, M. and M. Woodford, 1997, ``Rational Asset Pricing Bubbles'',
\textit{Econometrica} 65, 19--57.

\item Simon, H. A., 1955, ``A behavioral model of rational choice'', \textit{Quarterly Journal of Economics} 49,
99--118.
\end{description}

\end{document}